\documentstyle[nato]{crckapb} 

\begin{opening}
\title{Jets from black hole binaries and Galactic Nuclei}
\\

\author{I.F. Mirabel}
\institute{Centre d'Etudes de Saclay/ CEA/DSM/DAPNIA/SAP\\
           91911 Gif/Yvette, France \& \\
           Intituto de Astronom\'\i a y F\'\i sica del Espacio. Bs As, Argentina}
\end{opening}

\runningtitle{Jets from black holes}

\begin{document}

\section{Abstract}
Relativistic outflows are a common phenomenon in  
accreting black holes. Despite the enormous differences in scale, stellar-mass black holes 
in binaries and supermassive black holes in Galactic Nuclei produce jets with analogous 
properties. In both are observed two types of relativistic outflows: 
1) steady compact jets with flat-spectrum, 
and 2) sporadic extended jets with steep-spectrum and apparent superluminal motions. 
Besides, the most common class of gamma-ray bursts are afterglows from 
ultra-relativistic jets associated to the formation of black holes at cosmological distances. 

\vskip .1in

\section{Origin of the idea of stellar-mass and supermassive black hole}

Black holes were first predicted by John Michell (1784 ) in the context of Newtonian physics and 
the corpuscular theory of light. He proposed that these objects could be detected in binary 
systems by the 
motion of nearby luminous objects. In the fourth year of the french revolution, Pierre-Simon 
Laplace (1795) went even further speculating on the possible existence of both, stellar-mass 
and supermassive black holes. In the fourth edition of ``Exposition du Syt\`eme du Monde" 
Laplace suggested: 1) that stellar-mass black holes could be as numerous as 
stars -``en aussi grand nombre que les \'etoiles"-, and 
2) that the most massive objects of the universe could be black holes-``il est donc possible que 
les plus grands...corps de l'univers, soient par cela m\^eme, invisibles". In the XIXth 
century the ondulatory conception of light became predominant and the idea of black hole was 
forgotten for more than one hundred years, until it became a natural consequence of the 
general relativity conception of gravitation as a curvature of space-time. 

\begin{figure}
\vspace{17cm}
\caption{Diagram illustrating current ideas concerning quasars and microquasars (not to scale). As in quasars, in microquasars are found the following three basic ingredients: 1) a spinning black hole, 2) an accretion disk heated by viscous dissipation, and 3) collimated jets of relativistic particles. However, in microquasars the black hole is only a few solar masses instead of several millon solar masses; the accretion 
disk has mean thermal temperatures of several millon degrees instead of 
several thousand degrees; and the particles ejected at relativistic speeds 
can travel up to distances of a few light-years only, instead of the several millon light-years as in some giant radio galaxies. In quasars matter can be drawn into the accretion disk from disrupted stars or from the interstellar medium of the host galaxy, whereas in microquasars the material is being drawn from the companion star in the binary system. In quasars the accretion disk has sizes of $\sim$10$^9$ km and radiates mostly in the ultraviolet and optical wavelenghts, whereas in microquasars the accretion disk has sizes of $\sim$10$^3$ km and the bulk of the radiation comes out in the X-rays. 
It is believed that part of the spin energy of the black hole can be tapped to power the collimated ejection of magnetized plasma at relativistic speeds. This analogy between quasars and microquasars resides in the fact that in black holes the physics is essentially the same irrespective of the mass, except that the linear and time scales of phenomena are proportional 
to the black hole mass. Because of the relative proximity and shorter time scales, in microquasars it is possible to firmly establish the relativistic motion of the sources of radiation, and to better study the physics of accretion flows and jet formation near the horizon of black holes.} \label{fig-1}
\end{figure}

\section{The quasar-microquasar analogy}

The idea that supermassive black holes should exist at the centre of galaxies 
received strong support when the redshift of a 
quasi-stellar-radio-source ({\it quasar} was measured for the first time by 
Marteen Schmidt (1963). The cosmological distance of 
quasars indicated that they must be powered by the release of gravitational energy 
around supermassive compact objects, rather than nuclear fusion in the interior of stars. 
On the other hand, the first observational evidence for the existence of stellar-mass 
black holes came from the discovery of X-ray sources beyond the solar system by Riccardo   
Giacconi et al. (1962). The accurate X-ray location of Cygnus X-1 lead to a precise radio position, 
which allowed the optical identification of the black-hole binary. 

The recent finding in our own galaxy of {\it microquasars} (Margon, 1994; Mirabel et al. 1992) 
with apparent superluminal motions (Mirabel \& Rodr\'\i guez, 1994)   
has opened new perspectives for the astrophysics of black holes (Mirabel \& Rodr\'\i guez, 
1999 for a review). These scaled-down versions of quasars are believed to be 
powered by spinning 
black holes with masses of up to a few tens that of the Sun (see Figure 1). 
The word {\it microquasar} was 
chosen to suggest that the analogy with quasars is more than morphological, 
and that there is an underlying unity in the physics of accreting black 
holes over an enormous range of scales, from stellar-mass black holes in 
binary stellar systems, to supermassive black holes at the centre of distant 
galaxies (Rees, 1998).

At first glance it may seem paradoxical that relativistic jets were 
first discovered in the nuclei of galaxies and distant quasars and that 
for more than a decade SS433 was the only known object of its class in 
our Galaxy (Margon 1984). The reason for this is that disks 
around supermassive black holes emit strongly at optical and UV wavelengths.
Indeed, the more massive
the black hole, the cooler the surrounding accretion disk is.
For a black hole accreting at the Eddington limit,
the characteristic black body  
temperature at the last stable orbit in the
surrounding accretion disk will
be given approximately by $T \sim 2 \times 10^7~M^{-1/4}$
(Rees 1984), with
$T$ in K and the mass of the black hole, $M$, in solar masses.
Then, while accretion disks in AGNs have strong emission in the 
optical and ultraviolet with distinct 
broad emission lines, black hole and neutron star 
binaries usually are identified for the first time by their X-ray emission. 
Among these sources, SS 433 is unusual given its broad optical emission lines
and its brightness in the visible. 
Therefore, it is understandable that 
there was an impasse in the discovery of new stellar sources of relativistic 
jets until the recent developments in X-ray astronomy. 
Strictly speaking and if it had not been for the historical circumstances 
described above, the acronym  {\it quasar}  
would have suited better the stellar mass versions rather than their 
super-massive analogs at the centers of galaxies.

Since
the characteristic times in the flow of matter onto a black hole are proportional to its mass, variations with 
intervals of minutes in a microquasar correspond
to analogous phenomena with durations of thousands of years in a quasar of 
10$^9$ M$_{\odot}$, which is much longer than a human life-time Sams et al. 1996).
Therefore, variations with minutes of duration in microquasars could be sampling phenomena that we have not been able to study in quasars. 
The repeated observation of two-sided moving jets in a microquasar (Rodr\'\i guez \& Mirabel, 1999) has led to a much greater acceptance of the idea that the emission from quasar jets is associated with moving material at speeds close to that of light. Furthermore, simultaneous multiwavelength observations of this microquasar are revealing the connection between the sudden disappearance of matter through the horizon of the black hole, with the ejection of expanding clouds of relativistic plasma 
(see Figure 2). 

\begin{figure}
\vspace{4cm}
\caption{Radio, infrared, and X-ray light curves
for GRS~1915+105 at the time of quasi-periodic
oscillations on
1997 September 9 (Mirabel et al. 1998).
The infrared flare starts during the recovery from the X-ray dip,
when a sharp, isolated X-ray spike is observed.
These observations show the connection between the rapid disappearance
and follow-up replenishment of the inner accretion disk seen in the
X-rays (Belloni et al. 1997), and the ejection of relativistic plasma clouds 
observed as synchrotron 
emission at infrared wavelengths first and later at radio wavelengths.
A scheme of the relative positions where the different emissions originate
is shown in the top part of the figure.
The hardness ratio
(13-60 keV)/(2-13 keV) is shown at the bottom of the figure.} \label{fig-2}
\end{figure}

\section{Compact jets in stellar-mass and supermassive black holes}

The class of stellar-mass black holes that are persistent X-ray sources 
(e.g. Cygnus X-1, 1E 1740-2942, GRS 1758-258, etc.) and some 
supermassive black holes at the centre of galaxies (e.g. Sgr A$^*$ and many AGNs) do not exhibit  
luminous outbursts with large-scale sporadic ejections. However, despite  
the enormous differences in mass, steadily accreting black holes have analogous 
radio cores with steady, 
flat (S$_{\nu}$$\propto$$\nu$$^{\alpha}$; $\alpha$$\sim$0) emission at 
radio wavelengths. The fluxes of the core component in AGNs 
are typically of a few Janskys (e.g. Sgr A$^*$$\sim$1Jy) allowing VLBI high 
resolution studies, but in stellar mass black holes the cores are much fainter, typically of 
less than a few mJy, which makes difficult high resolution observations of the core.

Although there have been multiwavelength studies and speculation about the nature of the 
faint and steady compact radio emission in X-ray black hole 
binaries (e.g. Rodr\'\i guez et al. 1995; Fender et al. 1999, 2000), 
GRS 1915+105 is the black hole binary where the core 
has been succesfully imaged at AU scale resolution (Dhawan, Mirabel \& Rodr\'\i guez, 2000). 
GRS 1915+105 is the only X-ray binary where both, a compact core with 
steady fluxes $\geq$20 mJy, as well as large-scale superluminal ejections are 
{\it unambigously} observed. 
VLBA images during different states of the source  
always show compact jets with sizes $\sim$10$\lambda$$_{cm}$ AU along the same position angle 
as the superluminal large-scale jets (see Figure 3). As in the radio cores of AGNs, the 
brightness temperature of the compact jet in GRS 1915+105 is 
T$_B$$\geq$10$^9$ K. The VLBA images of GRS 
1915+105 are consistent with the conventional model of a conical expanding jet with syncrotron 
emission (Hjellming \& Johnston, 1988; Falke \& Biermann, 1999) in an optically thick region 
of solar system size.

\begin{figure}
\vspace{4cm}
\caption{Core of GRS 1915+105 resolved as a compact jet by 2-cm observations with the VLBA 
(Dhawan, Mirabel \& Rodr\'\i guez, 2000). At the 12 kpc 
distance of the source the angular resolution corresponds to 10 AU. Compact jets as 
are observed during quiescent, as well as QPO states as the one shown 
in Figure 2. The length of the compact jet and the period of the oscillations are 
consistent with bulk motions $\geq$0.9c, comparable with the velocities of the large-scale 
superluminal ejecta (Mirabel \& Rodr\'\i guez, 1994).} \label{fig-3}
\end{figure}

\section{Microblazars and gamma-ray bursts}

In all three galactic microquasars  
where $\theta$ (the angle 
between the line of sight and the axis of ejection) has been determined,
a large value is found (that is, the axis of ejection is close to the
plane of the sky). This result is not inconsistent with
the statistical expectation since the probability of
finding a source with a given $\theta$ 
is proportional to $sin~\theta$. We then expect to find as many
objects in the $60^\circ \leq \theta \leq 90^\circ$ range
as in the $0^\circ \leq \theta \leq 60^\circ$ range.
However, this argument suggests that we should eventually detect
objects with a small $\theta$. For objects with $\theta \leq 10^\circ$
we expect the timescales to be shortened by
2$\gamma$$^2$ and the flux densities to be boosted by 8$\gamma^3$ with respect to  
the values in the rest frame of the condensation.
For instance, for motions with $v$ = 0.98c ($\gamma$ = 5), the timescale will shorten
by a factor of $\sim$50 and the flux densities will be boosted by
a factor of $\sim 10^3$. Then, for a galactic source with
relativistic jets and small $\theta$ we expect fast and intense
variations in the observed flux.
These microblazars may be quite hard to detect
in practice, both because of the low probability
of small $\theta$ values and because of the fast decline
in the flux.

There is increasing evidence that the central engine of the most 
common form of gamma-ray burst (GRBs), those that last longer than 
a few seconds, are afterglows from ultra-relativistic jets produced during 
the formation of black holes (McFaden \& Woosley, 1999). Mirabel \& Rodr\'\i guez (1999) 
propose that ultra-relativistic 
bulk motion and beaming are needed to explain: 
1) the enormous energy requirements of $\geq$ 10$^{54}$ erg if the emission were 
isotropic (e.g. Kulkarni et al. 1999; 
Castro-Tirado et al. 1999); 2) the statistical correlation between time variability 
and brightness (Ramirez-Ruiz \& Fenimore, in Vth Compton workshop on GRBs 1999), 
and 3) the statistical anticorrelation 
between brightness and time-lag between hard and soft components (Norris et al. 1999). 
Beaming reduces 
the energy release by the beaming factor f = $\Delta$$\Omega$/4$\pi$, where 
$\Delta$$\Omega$ is the solid angle of the beamed emission. 
Additionally, the photon energies can be boosted to higher
values.
Extreme flows from collapsars with bulk 
Lorentz factors $>$ 100 have been proposed 
as sources of $\gamma$-ray bursts (M\'esz\'aros \& Rees 1997). High 
collimation (Dar 1998; 
Pugliese et al. 1999) can be tested observationaly (Rhoads, 1997), since 
the statistical properties of the bursts 
will depend on the viewing angle relative to the jet axis. 

Recent multiwavelength studies of gamma-ray afterglows suggest that they are highly 
collimated jets. The brightness of the optical transient 
associated to GRB 990123 showed a break (Kulkarni et al. 1999),  
and a steepening from a power law in time t proportional to 
t$^{-1.2}$, ultimately approaching 
a slope t$^{-2.5}$ (Castro-Tirado et al. 1999). The achromatic 
steepening of the optical light curve and early radio flux decay of 
GRB 990510 are inconsistent with simple spherical expansion, and 
well fit by jet evolution. 
It is interesting that 
the power laws that 
describe the light curves of the ejecta in microquasars 
show similar breaks and steepening of the 
radio flux density (Rodr\'\i guez \& Mirabel, 1999). 
In microquasars, these breaks and steepenings have been 
interpreted (Hjellming \& 
Johnston 1988) as a transition from slow intrinsic expansion followed 
by free expansion in two dimensions. Besides, linear polarizations of 
about 2\% were recently measured in the optical afterglow of 
GRB 990510 (Covino et al. 1999), 
providing strong evidence that the afterglow radiation from gamma-ray 
bursters is, at least in part, produced by synchrotron processes.  
Linear polarizations in the range of 2-10\% have been measured  
in microquasars at radio (e.g. Rodr\'\i guez et al. 1995), and optical 
(Scaltriti et al. 1997) wavelengths.

In this context, the jets in microquasars of our own Galaxy seem to be less extreme 
local analogs of the super-relativistic jets associated to the more 
distant gamma-ray bursters. However, there are caveats to this analogy and  
gamma-ray bursters are different to the 
microquasars found so far in our own Galaxy. The former do not repeat, 
seem to be related to catastrophic 
events, and have much larger super-Eddington luminosities. Therefore, 
the scaling laws in terms of the black hole mass that are valid in the 
analogy between microquasars and quasars do not seem to apply in the case of gamma-ray bursters.

\end{document}